\documentclass[]{article}
\usepackage{subfig}
\usepackage{graphicx}
\usepackage{float}
\usepackage[utf8]{inputenc}
\usepackage[margin=1in]{geometry}
\usepackage{authblk}

\begin{document}

\title{Graphlets versus node2vec and struc2vec in the task of network alignment}
\author[1, 2, 3]{Shawn Gu}
\author[1, 2, 3, *]{Tijana Milenkovi\'c}
\affil[1]{University of Notre Dame, Department of Computer Science and Engineering}
\affil[2]{Eck Institute for Global Health}
\affil[3]{Interdisciplinary Center for Network Science and Applications (iCeNSA)}
\affil[*]{To whom correspondence should be addressed}



\date{}
\maketitle

\begin{abstract}
    Network embedding aims to represent each node in a network as a low-dimensional feature vector that summarizes the given node's (extended) network neighborhood. The nodes' feature vectors can then be used in various downstream machine learning tasks. Recently, many embedding methods that automatically learn the features of nodes have emerged, such as node2vec and struc2vec, which have been used in tasks such as node classification, link prediction, and node clustering, mainly in the social network domain. There are also other embedding methods that explicitly look at the connections between nodes, i.e., the nodes' network neighborhoods, such as graphlets. Graphlets have been used in many tasks such as network comparison, link prediction, and network clustering, mainly in the computational biology domain. Even though the two types of embedding methods (node2vec/struct2vec versus graphlets) have a similar goal -- to represent nodes as features vectors, no comparisons have been made between them, possibly because they have originated in the different domains. Therefore, in this study, we compare graphlets to node2vec and struc2vec, and we do so in the task of network alignment. In evaluations on synthetic and real-world biological networks, we find that graphlets are both more accurate and faster than node2vec and struc2vec. 
    
    
\end{abstract}
\section{Introduction}

Many complex systems can be modeled as networks \cite{barabasi2016network, newman2010networks}. For example, social interactions between people can be modeled as social networks, and biochemical interactions between proteins inside the cell can be modeled as protein-protein interaction (PPI) networks. Modeling a system as a network allows us to consider the important interactions between entities (e.g., people, proteins, etc.), which can lead to deeper insights compared to analyzing each entity on its own. 

An important and popular computational problem in the field of network science is network embedding \cite{cui2017survey}. The goal of network embedding is to represent each node in a network as a low-dimensional feature vector such that the network structure is preserved. The nodes' feature vectors can then be used in various downstream machine learning tasks. For example, in the task of node classification, given a network where labels are known only for some of the nodes, one can embed all nodes in a low-dimensional space and train a classifier to predict labels of the other nodes based on their feature vector similarities, i.e., closeness in the space, to the labeled nodes \cite{grover2016node2vec, ribeiro2017struc2vec, dong2017metapath2vec}. In the task of link prediction, after obtaining a feature vector for each node, one can calculate  similarities between all node pairs, and then nodes with higher similarities will have higher probabilities of being linked \cite{yang2015evaluating, hulovatyy2014revealing}. In the task of network clustering, the nodes' (or edges') feature vectors can be given as input to a clustering algorithm to group similar nodes (or edges) together \cite{crawford2018cluenet,solava2012graphlet,milenkovic2008,malliaros2013clustering, harenberg2014community}.

Many network embedding methods \textit{automatically learn} the features of nodes. That is, these methods formulate the problem of embedding as the optimization of some objective function. Two recent state-of-the-art methods, \emph{node2vec} \cite{grover2016node2vec} and \emph{struc2vec}\cite{ribeiro2017struc2vec}, fall under this category. Intuitively, they use random walks to explore the extended neighborhood of a node and summarize it into the node's feature vector. To date, these approaches have been used in node classification, link prediction, and node clustering tasks, mainly in the social network domain \cite{grover2016node2vec, ribeiro2017struc2vec, dong2017metapath2vec, yang2015evaluating, malliaros2013clustering, harenberg2014community}. 

\begin{figure}[h!]
    \centering
 \includegraphics[width=0.48\textwidth]{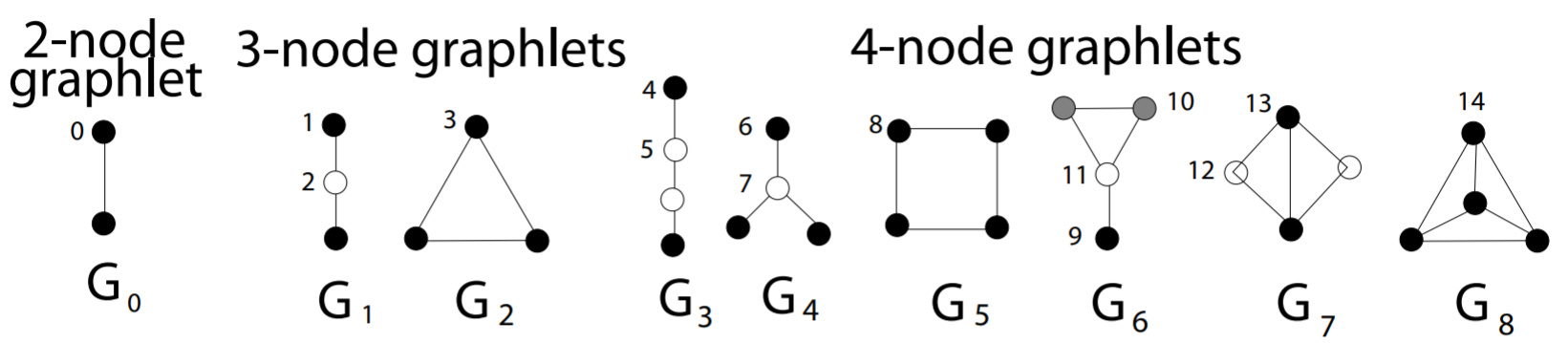}
    \caption{\label{fig:graphlets}Illustration of all nine 2-4-node graphlets and their 15 node symmetry groups (i.e., automorphism orbits). In a given graphlet, nodes belonging to the same orbit are colored the same.}   
\end{figure}

Other embedding methods \emph{explicitly look} at the  connections between nodes, i.e., the nodes' network neighborhoods, rather than trying to infer some features automatically through optimization. \emph{Graphlets} fall under this category. Graphlets (Fig. \ref{fig:graphlets}) are Lego-like building blocks of complex networks, i.e., small subgraphs of a network (a path, triangle, square, etc.). Graphlets can be used to summarize the extended neighborhood of a node into a feature vector as follows. For each node, for each topological node symmetry group (formally, automorphism orbit), one can count how many times the given node touches each graphlet at each of its orbits. The resulting counts for all graphlets/orbits form the node's \textit{graphlet degree vector (GDV)} \cite{milenkovic2008}. Graphlets and nodes' (as well as edges') GDVs have been used extensively in many tasks, such as network comparison, network clustering, and link prediction, mainly in the computational biology (i.e., biological network) domain \cite{milenkovic2008, yaverouglu2015proper, suncrawfordtangmilenkovic2015, hulovatyy2014revealing, solava2012graphlet, faisal2014dynamic, wang2014identification, singh2014graphlet}. 

Even though the methods that automatically learn node embeddings have a similar goal as graphlets -- to obtain a feature vector of a node -- to our knowledge no comparisons have been made between them, possibly because the two approach types have originated and have been used in the different domains (social versus biological networks). Even though recent studies on automatic learning-based embedding have recognized that graphlets can be seen as an alternative method for embedding, they have not compared against graphlets in their evaluations \cite{zitnik2017predicting, dong2017metapath2vec}. To close this gap and merge the knowledge from the different domains, in this study we compare the two types of embedding methods to each other. Specifically, we compare graphlets to node2vec and struc2vec, and we do so in the task of network alignment (NA).

\begin{figure}[h!]
    \centering
   \includegraphics[width=0.4\textwidth]{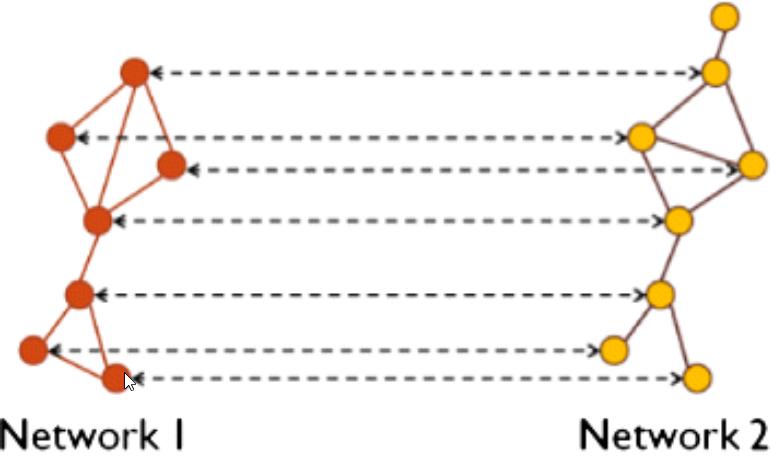}
    \caption{\label{fig:na-example} Illustration of network alignment (NA). NA aims to find a one-to-one (injective) mapping between nodes of the compared networks. Dotted lines represent nodes that are aligned to each other in this illustration. Note that NA as we define it in this study is called global NA, as it aims to  map all nodes from the smaller of the two networks to nodes from the larger network. Another type of NA exists called local NA, which typically results in only small (local) network regions being mapped to each other \cite{meng2016local,guzzi2017survey}, but this is out of the scope of our study. Also, note that NA as we define in this study is called pairwise NA, as it aims to align two networks. Another type of NA exists called multiple NA, which can align more than two networks \cite{vijayan2017multiple,vijayan2017pairwise}, but this is out of the scope of our study.}   
\end{figure}

Intuitively, NA aims to find a node mapping between the compared networks that uncovers network regions of high similarity (Fig. \ref{fig:na-example}) \cite{faisal2015post, meng2016local, guzzi2017survey}. This allows for the transfer of functional knowledge between the similar (i.e., aligned) network regions. NA is also referred to as alignment-based network comparison, graph matching, graph deanonymization, and identity matching. This is because NA has been applied to many domains. For example, if we align the PPI network of baker's yeast, a well-studied species, to the PPI network of human, a poorly-studied species, we can infer the function of human proteins based on the function of their aligned partners in the yeast network. This was done in order to study human aging \cite{faisal2015global}, which is otherwise difficult to do because of long life span and ethical constraints involving the human species. NA can also be used to deanonymize online social networks. For example, many Internet users are on multiple social media platforms, and thus, NA can be used to match identities (i.e., user accounts) across the different platforms \cite{zhang2015cosnet}. On the other hand, NA can be used to study how to prevent such deanonymization attacks on potentially sensitive data, thus having privacy implications \cite{narayanan2011link}. 

NA is computationally intractable, i.e., NP-hard, due to the NP-completeness of the underlying subgraph isomorphism problem \cite{cook1971complexity}. So, heuristic algorithms need to be sought. These algorithms typically consist of two parts. First, they measure pairwise similarities between nodes from the networks being aligned. Then, they use an alignment strategy to quickly identify alignments that maximize some objective function, which often takes into account the total similarity over all aligned nodes (i.e., the goal is to align similar nodes to each other), plus potentially the amount of edges that are conserved under the given node mapping. 

So, in our evaluation, we use each of graphlets (i.e., GDVs), node2vec, and struc2vec to quantify node similarities, plug those similarities into two established alignment strategies -- WAVE \cite{sun2015wave} and SANA \cite{sana}, and compare the results of the different node similarity measures under the same alignment strategy. We choose these two alignment strategies because they are recent and state-of-the-art methods, and also they are complementary to each other  in the sense that they use different algorithmic paradigms (WAVE uses a seed-and-extend approach while SANA uses a search algorithm) \cite{gu2017homogeneous}. We analyze synthetic  networks originating from different random graph models as well as real-world biological networks. We align each of the networks to its randomly rewired (i.e., noisy) versions and test the robustness of each NA approach to noise in the data. Since the aligned networks have only a percentage of their edges different but they have the same nodes, we know which nodes should be mapped to which. So, we quantify NA quality by measuring node correctness -- the percentage of nodes in the given alignment that are correctly mapped.

Because node2vec and struc2vec are based on random walks, and because random walks can be thought of as ``sampled graphlets'', we expect that node2vec and struc2vec will be faster than graphlets. However, we find that this is \emph{not} the case -- graphlets are an order of magnitude faster than both node2vec and struc2vec, even though the implementations of the latter two are parallelized, unlike the implementation of graphlets that we use. At the same time, graphlets overall yield more accurate alignments of the analyzed networks compared to both node2vec and struc2vec. Hence, our results imply that graphlets should be adopted from the computational biology domain to the social network domain.

 \section{Methods}

\subsection{Measuring node similarities}\label{sect:methods_node_similarities}

    \noindent \textbf{Graphlets}. Here, we provide more intuition behind graphlet automorphism orbits. Given a graphlet, nodes that are topologically symmetric to each other are a part of the same orbit. Consider a three node path ($G_1$ in Fig. \ref{fig:graphlets}). The two nodes on the end of the path are symmetric to each other, so they are in the same orbit (colored black and labeled with a $1$ in $G_1$). On the other hand, the node in the middle is only symmetric to itself, and thus in its own distinct orbit (colored white and labeled with a $2$ in $G_1$). So, if node $u$ in a network is on the end of a three node path, it ``participates" in the first orbit. If it is in the middle of a three node path, it ``participates" in the second orbit. These orbits are precomputed for all graphlets of up to five nodes. So, to form a node's GDV, for each orbit, one counts how many times that node participates in the orbit. Each position in the GDV corresponds to an orbit, so if we consider up to 4-node graphlets (which we do in this study), we have a feature vector of length 15, because 2-4-node graphlets have 15 orbits (Fig. \ref{fig:graphlets}). To compute the GDVs, we use the Orca tool \cite{hovcevar2014combinatorial}. 
    
    \vspace{2pt}
    \noindent \textbf{Node2vec}. Node2vec uses biased random walks (intuitively, these random walks follow the breadth-first search and depth-first search strategies) to learn the features of a node. In a random walk with some starting node $u$, in the first step there is an equal chance of visiting every neighbor of $u$. In the second step, starting from this neighbor of $u$, there is again an equal chance of visiting every neighbor. This continues until a specified number of steps is made. In a \textit{biased} random walk, in each step some neighbors have a higher or lower chance of being visited.
    
    In node2vec, given a node $u$, the algorithm performs a number of biased random walks of a certain length starting at $u$. Then, for every other node in the network, these walks are used to obtain the probability of the node $u$ being close in distance to the other node. For example, if many random walks starting at $u$ encounter some other node $v$ within a few steps, then there is a high probability $u$ is close to $v$. This also means the feature vector for $u$ will be similar to that of $v$. For the formal approach of node2vec, see \cite{grover2016node2vec}. 
    
    We use the default parameters of the C++ implementation provided in the SNAP GitHub; these parameters were chosen empirically in the study. 
    
    
    \vspace{2pt}
    \noindent \textbf{\textit{Struc2vec.}} Ribeiro \textit{el al.} introduce struc2vec as another biased random walk-based method. The main difference between node2vec and struc2vec is that while in node2vec random walks occur on the original network, in struc2vec they occur on a modified version of the network where nodes that are close in distance in this network are structurally similar in the original network. Using the node2vec example above, if node $u$ encounters some other node $v$ within a few steps in many random walks, there is a high probability $u$ is close, and therefore structurally similar, to $v$. Again, the feature vector of $u$ will be similar to that of $v$. For the formal approach of struc2vec, see \cite{ribeiro2017struc2vec}. 
    
    Ribeiro \textit{el al.} argue that while methods like node2vec work well for node classification tasks, they tend to fail for structural equivalence tasks. In particular, given a node with a certain feature, neighbors of the node are likely to have the same feature. As such, nodes that are close in distance will more likely have similar feature vectors compared to nodes that are far in distance. So, structural equivalence will not necessarily be captured very well.
    
    We use the default parameters provided in the struc2vec GitHub; these parameters were chosen empirically in the study. 
    

    
    \vspace{2pt}
    \noindent \textbf{Quantifying node similarity between networks.} Given two nodes and their respective feature vectors, we calculate the cosine similarity between them, which we then normalize between 0 and 1 by adding the maximum possible value (1) and dividing by the range (2). While the inverse of Euclidean distance is also an option for a similarity measure, empirically we have found that cosine similarity typically gives better results. Note that for graphlets, we first perform principal component analysis (PCA), a standard dimension reduction technique, on the GDVs of all nodes from the networks being aligned, and then we compute cosine similarity between the PCA-reduced GDVs. We use PCA because empirically we have found doing so gives better results. We choose the first $r$ principal components, where $r$ is at least two and as small as possible such that the $r$ components account for at least 90\% of the variation in the data. 
    
    \subsection{Alignment strategies}\label{sect:methods_al_strategies}
    
    \noindent \textbf{WAVE.} WAVE takes as input two networks and similarities between all pairs of nodes from the two networks. Then, it uses a ``seed-and-extend" algorithm to align the networks. First two highly similar nodes are aligned, i.e., seeded. Then, the seed's neighboring nodes that are similar are aligned, the seed's neighbor's neighbors that are similar are aligned, and so on. The extension step continues until all nodes in the smaller of the two compared networks are aligned (formally, until a one-to-one node mapping between the two networks is produced). 
    
    \vspace{2pt}
    \noindent \textbf{SANA.} SANA also takes as input two networks and similarities between all pairs of nodes from the two networks, like WAVE. However, SANA uses a search algorithm, specifically simulated annealing, to find an alignment. That is, instead of aligning networks node by node as WAVE does, SANA explores the space of possible alignments and find the highest scoring one with respect to the objective function. We set the following parameters for SANA: \texttt{s3} to 1, \texttt{esim} to 1, \texttt{simFile} to the name of the node similarity file, and \texttt{simFormat} to 1 (this tells SANA to read the similarity file such that each line has 3 columns: node1, node2, and the similarity between them). SANA also has a running time parameter; for the smaller synthetic networks (discussed below), we set \texttt{t} to 5 minutes. For the larger biological networks (discussed below) we increase \texttt{t} to 60 minutes since SANA requires more time to find a good alignment (which we have verified empirically in our evaluation).
    
    \section{Results and discussion}

    \subsection{Evaluation}

    We study the three embedding (i.e., node similarity) methods -- graphlets, node2vec, and struc2vec (Section \ref{sect:methods_node_similarities}). We run each embedding method under each of the two alignment strategies -- WAVE and SANA (Section \ref{sect:methods_al_strategies}). We evaluate the three embedding methods under the same alignment strategy by aligning two types of networks -- synthetic and real-world biological (PPI) networks, to their noisy counterparts, as follows.
    
    \vspace{2pt}
    \noindent \textbf{Synthetic networks.} We form synthetic networks using two random graph generators, namely: 1) geometric random graphs (GEO) and 2) scale-free networks (SF). The two models have distinct network topologies \cite{milenkovic2008graphcrunch}, which enables us to test the robustness of our results to the choice of random graph model. We set both networks to the same size of 1,000 nodes and 6,000 edges.
    
    \vspace{2pt}
    \noindent \textbf{PPI networks.} We consider two different types of PPIs (i.e., edges): only affinity capture coupled to mass spectrometry (APMS), and only two-hybrid (Y2H). Sizes of the two PPI networks are shown in Table \ref{tab:ppi-sizes}. The two edge types reflect biological experiments used to detect the PPIs. Intuitively, APMS experiments will result in networks that will have more clique-like structures, and Y2H experiments will result in networks that will have more star-like structures. That is, the two PPI networks have different topologies, which allows us to test the robustness of our results to the choice of PPI type.
    
    
\begin{table}[h]
\begin{center}
\begin{tabular}{ c c c }
 \hline
 Network & \# of nodes & \# of edges  \\
 \hline
 APMS & 11,450 & 92,257 \\ 
 Y2H & 10,317 & 41,925 \\  
 \hline
\end{tabular}
\caption{Sizes of the two considered PPI networks.}\label{tab:ppi-sizes}
\end{center}
\end{table}

    \vspace{2pt}
    \noindent \textbf{Creating noisy counterparts of a synthetic or PPI network.} A noisy counterpart is the original network with $x\%$ of its edges rewired, where we vary $x$ to be 0, 10, 25, 50, 75, and 100. Since only edges are changed between the original and noisy network, we know which nodes should be mapped to which. We can use this true node mapping to accurately evaluate our methods; a good method should have high node correctness, which is the percentage of node pairs from the given alignment that are present in the true node mapping. We form five rewired network instances at each noise level to account for the randomness of the edge rewiring process. Then, we average the alignment quality over the multiple runs corresponding to the multiple instances.

\subsection{Method comparison in terms of accuracy}

\begin{figure}[h!]
    \centering
\includegraphics[width=0.48\textwidth]{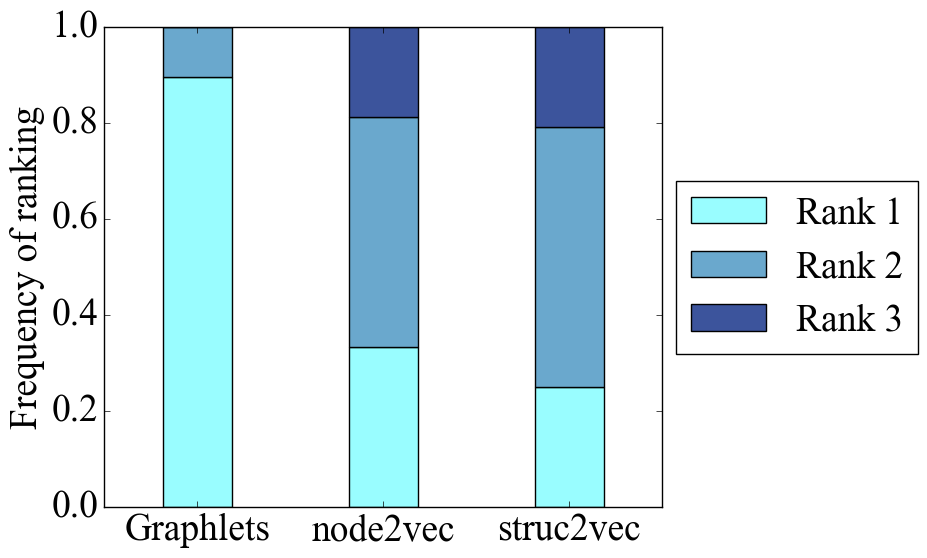}
    \caption{\label{fig:res} Summarized results regarding the effect of the embedding method ($x$-axis) on alignment quality over 32  evaluation tests (4 networks $\times$ 4 noise levels $\times$ 2 alignment strategies). For each test, we compare the different methods and rank them from the best (rank 1) to the worst (rank 3). The figure shows the percentage (frequency) of all evaluation tests in which the given method has the given rank. Note that in the figure, some rank ties exist (see the text for details). }   
    \end{figure}

\begin{figure*}[h!]
\centering
    \subfloat[]{\includegraphics[width=0.3\textwidth]{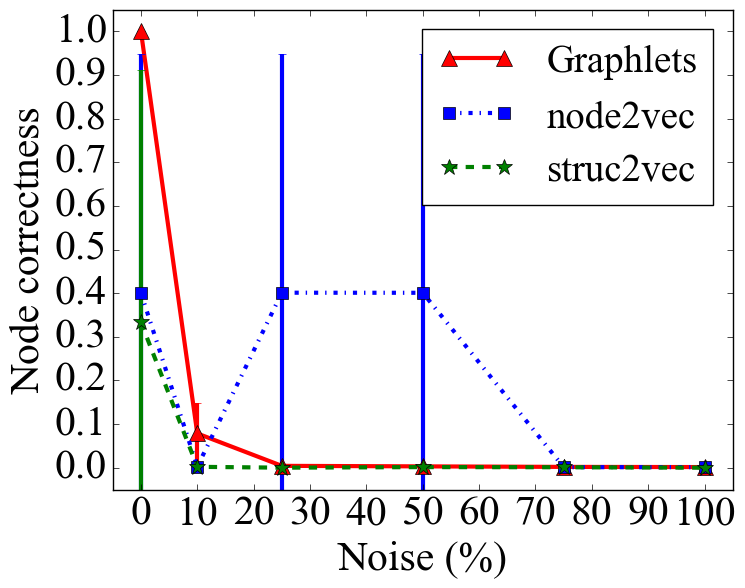}}
    \subfloat[]{\includegraphics[width=0.3\textwidth]{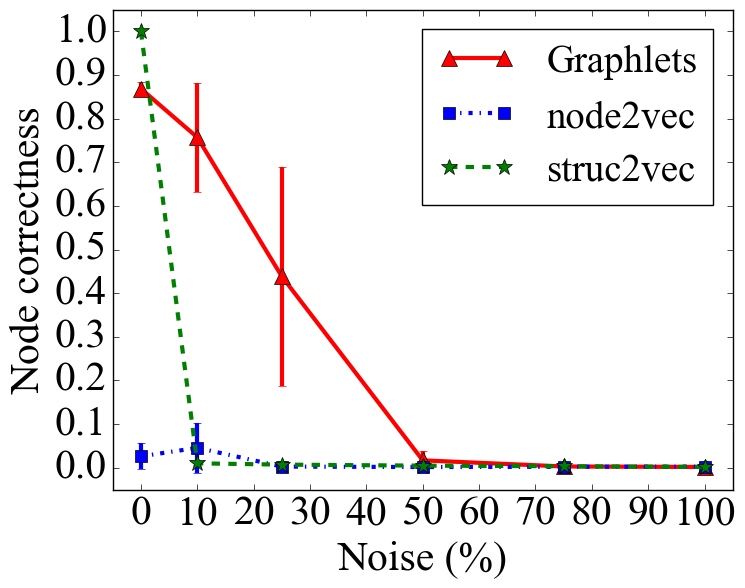}}\\
    \subfloat[]{\includegraphics[width=0.3\textwidth]{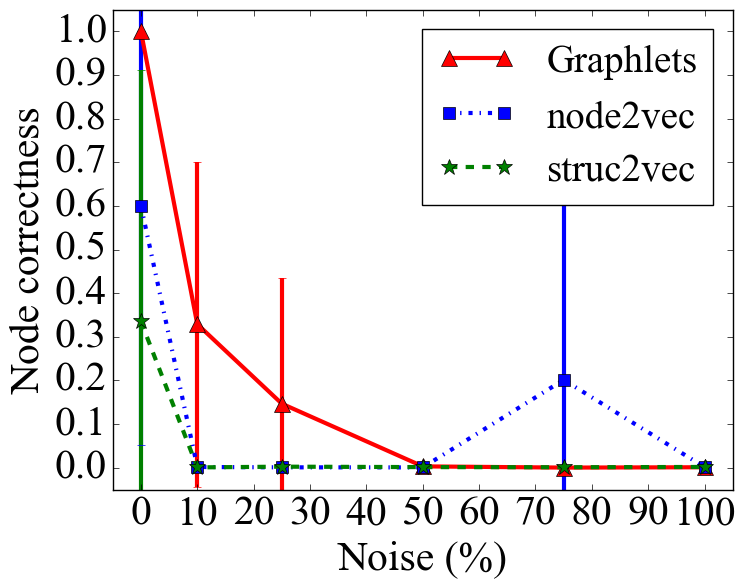}}
    \subfloat[]{\includegraphics[width=0.3\textwidth]{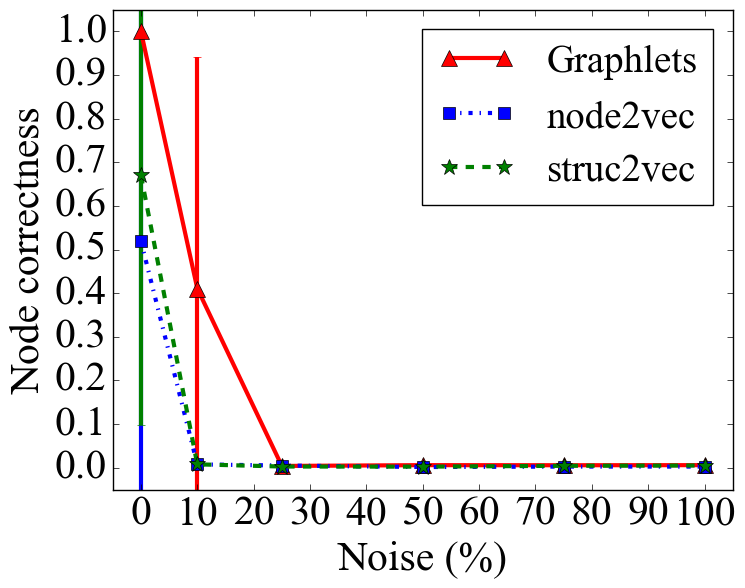}} \\
    \subfloat[]{\includegraphics[width=0.3\textwidth]{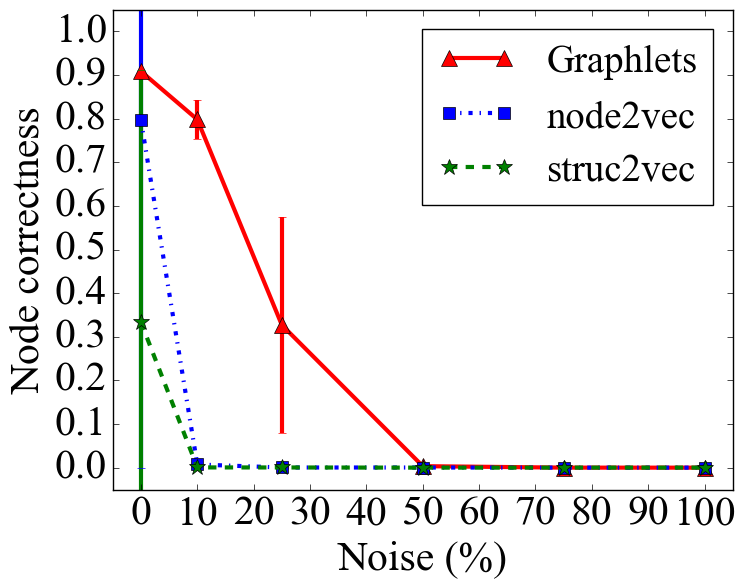}}
    \subfloat[]{\includegraphics[width=0.3\textwidth]{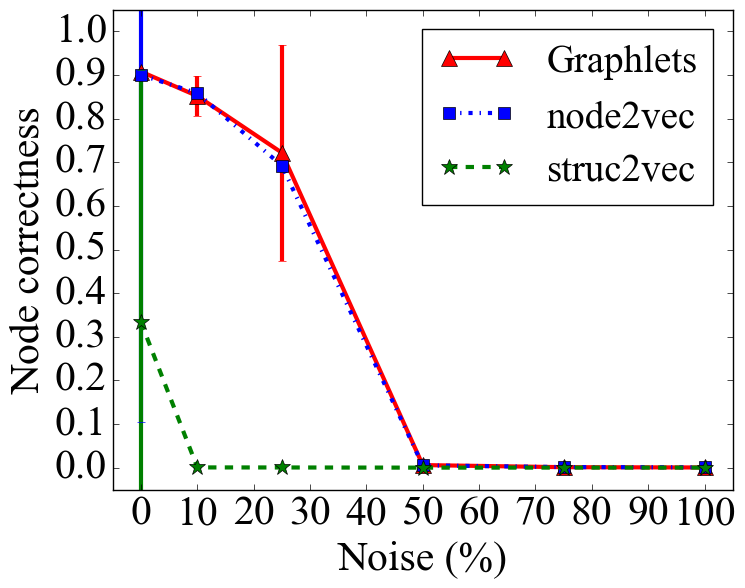}}\\
    \subfloat[]{\includegraphics[width=0.3\textwidth]{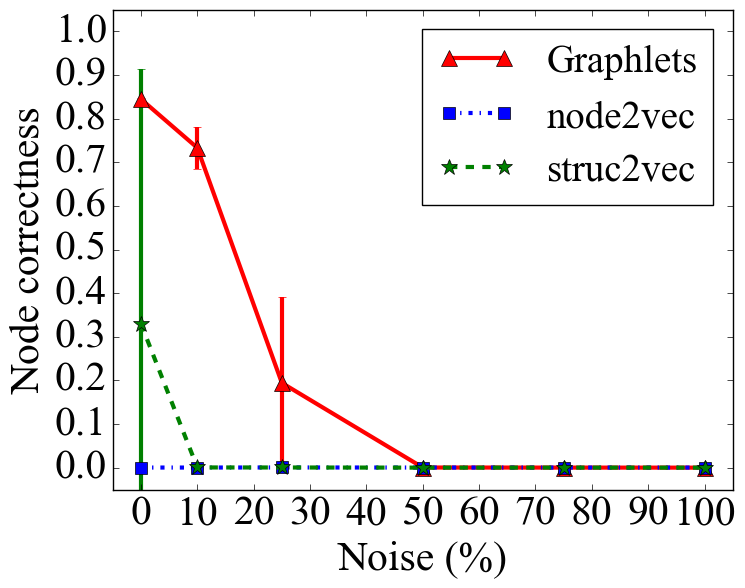}}
    \subfloat[]{\includegraphics[width=0.3\textwidth]{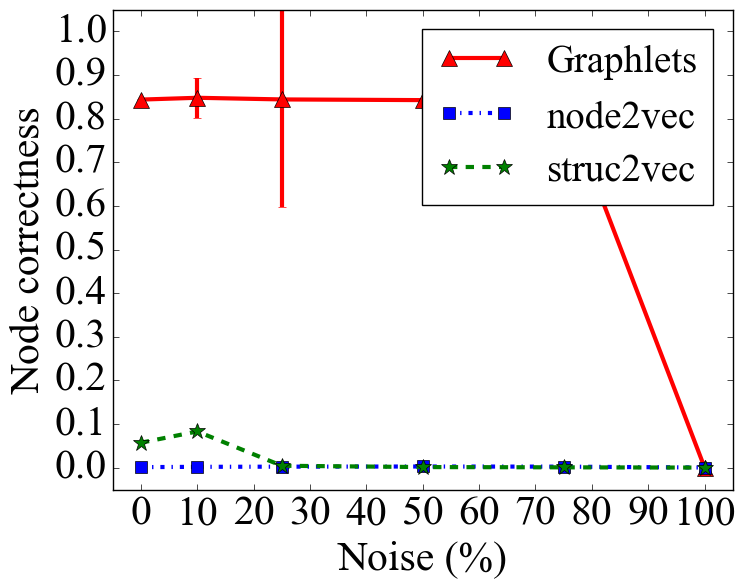}}
\caption{\label{fig:detailed_results} Detailed alignment quality results regarding the effect of the embedding method on alignment quality as a function of noise for (a,b) geometric synthetic networks, (c,d) scale-free synthetic networks, (e,f) the APMS PPI real-world network, and (g,h) the Y2H PPI real-world network, under (a,c,e,g) WAVE alignment strategy and (b,d,f,h) SANA alignment strategy.}   
\end{figure*}

    We expect that a good method (i.e., a combination of an embedding method and an alignment strategy) should have high alignment quality at low noise levels, and low alignment quality at high noise levels, with a general decreasing trend as noise increases. This is because at low noise levels, we are aligning two networks with similar topologies compared to each other, while at high noise levels, we are aligning two networks with almost random topologies compared to each other. We also expect that a good method should be robust to noise. That is, we should see a slower decrease in alignment quality as noise increases compared to other methods. 
    We find that graphlets fit all of these criteria (Figs. \ref{fig:res} and \ref{fig:detailed_results}), unlike node2vec or struc2vec.
    
    Specifically, we first perform a summary analysis that aims to rank the three embedding approaches against each other. Considering up to 50\% noise (which results in 4 networks $\times$ 4 noise levels $\times$ 2 alignment strategies $=$ 32 evaluation tests), we count how many times out of the total of 32 evaluation tests a given embedding method is the best (rank 1), second best (rank 2), or the third best, i.e., worst (rank 3) in terms of alignment quality. Note that in this ranking analysis we omit the largest noise levels of 75\% and 100\% because for most of the methods and networks, alignment quality is tied at near 0 values, as expected for such random-like noise levels.
    
    In the ranking analysis, we find that graphlets are the best approach (have rank 1) most of the time -- in 90.63\% of all evaluation tests (Fig. \ref{fig:res}). We do observe some ties between the methods. In particular, all three methods are tied with each other in 21.88\% of all cases, and graphlets are tied with node2vec (but not with struc2vec) in 6.25\% of additional cases. This means that graphlets are superior (without ties) to both node2vec and struc2vec in 62.5\% of all cases, and are tied to at least one of node2vec or struc2vec in additional  $21.88 \% + 6.25\% = 28.13\%$ of all cases. Node2vec is superior to the other two methods in only 6.25\% of all cases. Struc2vec is superior to the other two methods in only 3.125\% of all cases. This analysis questions the usefulness of node2vec and struc2vec, as they improve upon graphlets in only two and one out of the 32 cases, respectively.

    Second,  we consider the detailed alignment quality versus noise results for each embedding method, network, and alignment strategy. This detailed analysis again confirms the superiority of graphlets
    (Fig. \ref{fig:detailed_results}). Specifically, at low noise levels, graphlets have high alignment quality, and at high noise levels, graphlets have low alignment quality, with a decreasing trend, as expected. Contrast this with node2vec's results, where at some higher noise levels the alignment quality is actually better than at lower noise levels, which is a trend that should not happen, as the networks being aligned are more similar and should thus yield higher alignment quality at lower than at higher noise levels. Regarding struc2vec, while this method overall shows the expected trends, just as graphlets, it is less accurate than graphlets. Also, of all approaches, we find that graphlets are the most robust to noise, showing a slower decrease in alignment quality as noise increases compared to the other two methods. 
    
    


\subsection{Method comparison in terms of running time}

    We also analyze running times of the different embedding methods. All methods are run on a 64-core AMD Opteron 6376 machine with 500 GB of RAM.
    We record both real times (the actual clock time elapsed) and CPU times (the total amount of time the core(s) spend executing) for all methods. The more cores are used, the lower the real time is expected to be. Hence, it is more fair to compare the different methods' CPU times, which essentially reflect how long the given method would take if a single core was used. This is especially true because node2vec uses 10 cores, struc2vec uses four cores, and the implementation of graphlets that we consider uses only one core. Again, if one wishes to ignore implementation-specific (dis)advantages of the given method, including (lack of) parallelization, it is more fair to compare the different methods' CPU running times. On the other hand, if one wishes to give each method/implementation the best-case advantage, then the methods' real times should be compared.
    
    
    
    We expect computing node2vec and struc2vec to be faster than counting graphlets, because the former two are based on random walks, which can be thought of as ``sampled graphlets''. However, we find that counting graphlets, i.e., obtaining nodes' GDVs, is faster than computing the nodes'  node2vec or struc2vec feature vectors. This holds for all four analyzed networks, and independent on whether we consider real or CPU running times (Table \ref{tab:running-times}).  Also, we note that struc2vec is slower than node2vec. One possible explanation for this is that struc2vec has an extra step compared to node2vec -- that of creating a modified version of the original network when performing random walks.
    
    The superiority of graphlets in terms of running time is most likely due to the graphlet implementation that we use called Orca, which leverages combinatorial relationships between the different graphlet orbits. That is, by knowing the counts of some graphlets (i.e., orbits), Orca can infer the counts of the other graphlets through mathematical equations, rather than having to actually count these graphlet occurrences by traversing  the network structure. Consequently, Orca significantly speeds up computation of graphlet counts compared to the naive implementation of graphlet counting that would explicitly search the network structure for the occurrence of every graphlet, such as that implemented in the GraphCrunch tool \cite{milenkovic2008graphcrunch}.

    
    
\begin{table}[h!]
\centering
\caption{Running times of the embedding methods on each network, in seconds. In the table, real running time refers to the actual clock time elapsed, while CPU running time refers to the amount of time the core(s) spend executing.}
\label{tab:running-times}
\begin{tabular}{l | rrrr}
               & GEO              & SF               & APMS             & Y2H      \\
               \hline
graphlets-real & 0.03            & 0.03            & 1.08            & 0.26    \\
graphlets-CPU & 0.02 & 0.02 & 1.02 & 0.22 \\
node2vec-real  & 9.75 & 13.38           & 105.56          & 122.91  \\
node2vec-CPU   & 563.40          & 777.73 & 6426.68 & 7504.04 \\
struc2vec-real & 75.37 & 77.09           & 2874.24         & 1942.90 \\
struc2vec-CPU  & 268.03          & 282.43          & 10998.21        & 7333.74
\end{tabular}
\end{table}
    
    \section{Conclusion}
    
    In summary, we compare the three network embedding methods -- graphlets, node2vec, and struc2vec, in the task of network alignment. Specifically, for a given embedding method, we use the features generated by it to calculate node similarities. Then, we use the node similarity information in two existing network alignment strategies, WAVE and SANA. We fairly evaluate the different embedding methods under the same alignment strategy, and we do so by aligning synthetic and PPI networks to their noisy counterparts. We find that graphlets generally outperform the other embedding methods in terms of alignment quality. Importantly, node2vec and struc2vec improve upon graphlets in under 6.25\% of all  evaluation tests, which questions their usefuleness compared to graphlets, at least in the considered task of network alignment. Furthermore, not only do we find graphlet-based alignments to be the most accurate, but we also find that counting graphlets is faster than computing node2vec and struc2vec. We use the Orca implementation for graphlet counting. There  also exist more recent implementations than Orca that aim to  speed up graphlet counting,  e.g., by parallelizing the counting or estimating counts through sampling, which consequently makes graphlet counting possible in very large networks with millions of nodes/edges \cite{ortmann2016quad, wang2016fast, ahmed2015efficient}. These implementations could further speed up graphlet counting.
    
    Evaluating how graphlets perform against node2vec and struc2vec in tasks other than that of network alignment, such as node classification, network clustering, or link prediction, is the subject of future work.
    
    Importantly, while graphlets were originally introduced in the context of static and homogeneous networks \cite{prvzulj2004modeling,milenkovic2008}, given recent availability of dynamic (temporal, evolving) or heterogeneous (multi-node or multi-edge type) networks, graphlets have  been extended into their dynamic \cite{hulovatyy2015exploring} or heterogeneous \cite{gu2017homogeneous} counterparts. Then, dynamic or heterogenous graphlets have been used in newly defined network science tasks of aligning dynamic (rather than traditionally static) networks \cite{vijayan2017aligning, vijayan2017alignment},  clustering a dynamic network based on topological similarity rather than the traditionally considered notion of network denseness \cite{crawford2018cluenet}, and aligning heterogeneous (rather than traditionally homogeneous) networks \cite{gu2017homogeneous}. 
    
    Therefore, we believe that graphlets will continue to make their mark in the field of network science, hopefully across its many interdisciplinary domains, including biological, social, technological, information, transportation, infrastructure, ecological, climate, and other networks.
    
    \section*{Acknowledgements}

    This work was funded by Air Force Office of Scientific Research (AFOSR) Young Investigator Research Program (YIP) under award number FA9550-16-1-0147, and National Science Foundation (NSF) Faculty Early Career Development Program (CAREER) under award number CCF-1452795.


\begin{thebibliography}{10}

\bibitem{ahmed2015efficient}
N.~K. Ahmed, J.~Neville, R.~A. Rossi, and N.~Duffield.
\newblock Efficient graphlet counting for large networks.
\newblock In {\em Data Mining (ICDM), 2015 IEEE International Conference on},
  pages 1--10. IEEE, 2015.

\bibitem{barabasi2016network}
A.-L. Barab{\'a}si.
\newblock {\em Network science}.
\newblock Cambridge University Press, 2016.

\bibitem{cook1971complexity}
S.~A. Cook.
\newblock The complexity of theorem-proving procedures.
\newblock In {\em Proceedings of the third annual ACM Symposium on Theory of
  Computing}, pages 151--158. ACM, 1971.

\bibitem{crawford2018cluenet}
J.~Crawford and T.~Milenkovi{\'c}.
\newblock Cluenet: Clustering a temporal network based on topological
  similarity rather than denseness.
\newblock {\em PLOS ONE, in press}, 2018.

\bibitem{cui2017survey}
P.~Cui, X.~Wang, J.~Pei, and W.~Zhu.
\newblock A survey on network embedding.
\newblock {\em arXiv preprint arXiv:1711.08752}, 2017.

\bibitem{dong2017metapath2vec}
Y.~Dong, N.~V. Chawla, and A.~Swami.
\newblock metapath2vec: Scalable representation learning for heterogeneous
  networks.
\newblock In {\em Proceedings of the 23rd ACM SIGKDD International Conference
  on Knowledge Discovery and Data Mining}, pages 135--144. ACM, 2017.

\bibitem{faisal2015post}
F.~E. Faisal, L.~Meng, J.~Crawford, and T.~Milenkovi{\'c}.
\newblock The post-genomic era of biological network alignment.
\newblock {\em EURASIP Journal on Bioinformatics and Systems Biology},
  2015(1):3, 2015.

\bibitem{faisal2014dynamic}
F.~E. Faisal and T.~Milenkovi{\'c}.
\newblock Dynamic networks reveal key players in aging.
\newblock {\em Bioinformatics}, 30(12):1721--1729, 2014.

\bibitem{faisal2015global}
F.~E. Faisal, H.~Zhao, and T.~Milenkovi{\'c}.
\newblock Global network alignment in the context of aging.
\newblock {\em IEEE/ACM Transactions on Computational Biology and
  Bioinformatics}, 12(1):40--52, 2015.

\bibitem{grover2016node2vec}
A.~Grover and J.~Leskovec.
\newblock node2vec: Scalable feature learning for networks.
\newblock In {\em Proceedings of the 22nd ACM SIGKDD International Conference
  on Knowledge Discovery and Data Mining}, pages 855--864. ACM, 2016.

\bibitem{gu2017homogeneous}
S.~Gu, J.~Johnson, F.~E. Faisal, and T.~Milenkovi{\'c}.
\newblock From homogeneous to heterogeneous network alignment.
\newblock {\em arXiv preprint arXiv:1704.01221}, 2017.

\bibitem{guzzi2017survey}
P.~H. Guzzi and T.~Milenkovi{\'c}.
\newblock Survey of local and global biological network alignment: the need to
  reconcile the two sides of the same coin.
\newblock {\em Briefings in Bioinformatics}, 2017.

\bibitem{harenberg2014community}
S.~Harenberg, G.~Bello, L.~Gjeltema, S.~Ranshous, J.~Harlalka, R.~Seay,
  K.~Padmanabhan, and N.~Samatova.
\newblock Community detection in large-scale networks: a survey and empirical
  evaluation.
\newblock {\em Wiley Interdisciplinary Reviews: Computational Statistics},
  6(6):426--439, 2014.

\bibitem{hovcevar2014combinatorial}
T.~Ho{\v{c}}evar and J.~Dem{\v{s}}ar.
\newblock A combinatorial approach to graphlet counting.
\newblock {\em Bioinformatics}, 30(4):559--565, 2014.

\bibitem{hulovatyy2015exploring}
Y.~Hulovatyy, H.~Chen, and T.~Milenkovi{\'c}.
\newblock Exploring the structure and function of temporal networks with
  dynamic graphlets.
\newblock {\em Bioinformatics}, 31(12):i171--i180, 2015.

\bibitem{hulovatyy2014revealing}
Y.~Hulovatyy, R.~W. Solava, and T.~Milenkovi{\'c}.
\newblock Revealing missing parts of the interactome via link prediction.
\newblock {\em PLOS ONE}, 9(3):e90073, 2014.

\bibitem{malliaros2013clustering}
F.~D. Malliaros and M.~Vazirgiannis.
\newblock Clustering and community detection in directed networks: A survey.
\newblock {\em Physics Reports}, 533(4):95--142, 2013.

\bibitem{sana}
N.~Mamano and W.~B. Hayes.
\newblock Sana: simulated annealing far outperforms many other search
  algorithms for biological network alignment.
\newblock {\em Bioinformatics}, 33(14):2156--2164, 2017.

\bibitem{meng2016local}
L.~Meng, A.~Striegel, and T.~Milenkovi{\'c}.
\newblock Local versus global biological network alignment.
\newblock {\em Bioinformatics}, 32(20):3155--3164, 2016.

\bibitem{milenkovic2008}
T.~Milenkov{\'c} and N.~Pr{\v{z}}ulj.
\newblock Uncovering biological network function via graphlet degree
  signatures.
\newblock {\em Cancer Informatics}, 6, 2008.

\bibitem{milenkovic2008graphcrunch}
T.~Milenkovi{\'c}, J.~Lai, and N.~Pr{\v{z}}ulj.
\newblock Graphcrunch: a tool for large network analyses.
\newblock {\em BMC bioinformatics}, 9(1):70, 2008.

\bibitem{narayanan2011link}
A.~Narayanan, E.~Shi, and B.~I. Rubinstein.
\newblock Link prediction by de-anonymization: How we won the kaggle social
  network challenge.
\newblock In {\em Neural Networks (IJCNN), The 2011 International Joint
  Conference on}, pages 1825--1834. IEEE, 2011.

\bibitem{newman2010networks}
M.~Newman.
\newblock {\em Networks: an introduction}.
\newblock Oxford University Press, 2010.

\bibitem{ortmann2016quad}
M.~Ortmann and U.~Brandes.
\newblock Quad census computation: Simple, efficient, and orbit-aware.
\newblock In {\em International Conference and School on Network Science},
  pages 1--13. Springer, 2016.

\bibitem{prvzulj2004modeling}
N.~Pr{\v{z}}ulj, D.~G. Corneil, and I.~Jurisica.
\newblock Modeling interactome: scale-free or geometric?
\newblock {\em Bioinformatics}, 20(18):3508--3515, 2004.

\bibitem{ribeiro2017struc2vec}
L.~F. Ribeiro, P.~H. Saverese, and D.~R. Figueiredo.
\newblock struc2vec: Learning node representations from structural identity.
\newblock In {\em Proceedings of the 23rd ACM SIGKDD International Conference
  on Knowledge Discovery and Data Mining}, pages 385--394. ACM, 2017.

\bibitem{singh2014graphlet}
O.~Singh, K.~Sawariya, and P.~Aparoy.
\newblock Graphlet signature-based scoring method to estimate protein--ligand
  binding affinity.
\newblock {\em Royal Society Open Science}, 1(4):140306, 2014.

\bibitem{solava2012graphlet}
R.~W. Solava, R.~P. Michaels, and T.~Milenkovi{\'c}.
\newblock Graphlet-based edge clustering reveals pathogen-interacting proteins.
\newblock {\em Bioinformatics}, 28(18):i480--i486, 2012.

\bibitem{suncrawfordtangmilenkovic2015}
Y.~Sun, J.~Crawford, J.~Tang, and T.~Milenkovi{\'c}.
\newblock Simultaneous optimization of both node and edge conservation in
  network alignment via wave.
\newblock {\em Lecture Notes in Computer Science Algorithms in Bioinformatics},
  pages 16--39, 2015.

\bibitem{sun2015wave}
Y.~Sun, J.~Crawford, J.~Tang, and T.~Milenkovi{\'c}.
\newblock Simultaneous optimization of both node and edge conservation in
  network alignment via wave.
\newblock {\em Lecture Notes in Computer Science Algorithms in Bioinformatics},
  pages 16--39, 2015.

\bibitem{vijayan2017alignment}
V.~Vijayan, D.~Critchlow, and T.~Milenkovi{\'c}.
\newblock Alignment of dynamic networks.
\newblock {\em Bioinformatics}, 33(14):i180--i189, 2017.

\bibitem{vijayan2017pairwise}
V.~Vijayan, E.~Krebs, L.~Meng, and T.~Milenkovi{\'c}.
\newblock Pairwise versus multiple network alignment.
\newblock {\em arXiv preprint arXiv:1709.04564}, 2017.

\bibitem{vijayan2017aligning}
V.~Vijayan and T.~Milenkovi{\'c}.
\newblock Aligning dynamic networks with dynawave.
\newblock {\em Bioinformatics}, 2017.

\bibitem{vijayan2017multiple}
V.~Vijayan and T.~Milenkovi{\'c}.
\newblock Multiple network alignment via multimagna++.
\newblock {\em IEEE/ACM Transactions on Computational Biology and
  Bioinformatics}, 2017.

\bibitem{wang2016fast}
P.~Wang, X.~Zhang, Z.~Li, J.~Cheng, J.~Lui, D.~Towsley, J.~Zhao, J.~Tao, and
  X.~Guan.
\newblock A fast sampling method of exploring graphlet degrees of large
  directed and undirected graphs.
\newblock {\em arXiv preprint arXiv:1604.08691}, 2016.

\bibitem{wang2014identification}
X.-D. Wang, J.-L. Huang, L.~Yang, D.-Q. Wei, Y.-X. Qi, and Z.-L. Jiang.
\newblock Identification of human disease genes from interactome network using
  graphlet interaction.
\newblock {\em PLOS ONE}, 9(1):e86142, 2014.

\bibitem{yang2015evaluating}
Y.~Yang, R.~N. Lichtenwalter, and N.~V. Chawla.
\newblock Evaluating link prediction methods.
\newblock {\em Knowledge and Information Systems}, 45(3):751--782, 2015.

\bibitem{yaverouglu2015proper}
{\"O}.~N. Yavero{\u{g}}lu, T.~Milenkovi{\'c}, and N.~Pr{\v{z}}ulj.
\newblock Proper evaluation of alignment-free network comparison methods.
\newblock {\em Bioinformatics}, 31(16):2697--2704, 2015.

\bibitem{zhang2015cosnet}
Y.~Zhang, J.~Tang, Z.~Yang, J.~Pei, and P.~S. Yu.
\newblock Cosnet: Connecting heterogeneous social networks with local and
  global consistency.
\newblock In {\em Proceedings of the 21th ACM SIGKDD International Conference
  on Knowledge Discovery and Data Mining}, pages 1485--1494. ACM, 2015.

\bibitem{zitnik2017predicting}
M.~Zitnik and J.~Leskovec.
\newblock Predicting multicellular function through multi-layer tissue
  networks.
\newblock {\em Bioinformatics}, 33(14):i190--i198, 2017.

\end{thebibliography}
\end{document}